\documentclass[5p,twocolumn]{elsarticle}

\usepackage{hyperref}
\usepackage{amsmath}
\usepackage{amssymb}
\usepackage{caption}
\usepackage{color}
\usepackage{graphicx}
\usepackage[utf8]{inputenc}
\usepackage{multirow}
\usepackage{wasysym}
\usepackage{nidanfloat}
\usepackage{subcaption}
\usepackage[normalem]{ulem}
\usepackage{siunitx}
\usepackage{threeparttable} 

\usepackage{array}
\newcolumntype{C}[1]{>{\centering\arraybackslash}p{#1}}

\begin{document}

\begin{frontmatter}

\title{Constraining the $^{22}$Ne($\alpha$,$\gamma$)$^{26}$Mg and $^{22}$Ne($\alpha$,n)$^{25}$Mg reaction rates using sub-Coulomb $\alpha$-transfer reactions}

\author[1,2]{H. Jayatissa\corref{cor1}\fnref{fn1}}
\ead{hjayatissa@anl.gov}

\author[1,2,3]{G.V. Rogachev\corref{cor1}}
\ead{rogachev@tamu.edu}

\author[2]{V.Z. Goldberg}
\author[2]{E. Koshchiy}
\author[1,2,3]{G. Christian\fnref{fn2}}
\author[1,2]{J. Hooker\fnref{fn3}}
\author[2]{S. Ota}
\author[2]{B.T. Roeder}
\author[2]{A. Saastamoinen}
\author[4]{O. Trippella}
\author[1,2]{S. Upadhyayula}
\author[2]{E. Uberseder}

\cortext[cor1]{Corresponding authors}

\fntext[fn1] {Present address: Physics Division, Argonne National Laboratory, Argonne, IL 60439, USA}
\fntext[fn2] {Present address: Department of Astronomy \& Physics, Saint Mary's University, Halifax, NS B3H 3C3 Canada}
\fntext[fn3] {Present address: Department of Physics \& Astronomy, University of Tennessee, Knoxville, TN 37996, USA}

\address[1]{Department of Physics \& Astronomy, Texas A\&M University, College Station, TX 77843, USA}
\address[2]{Cyclotron Institute, Texas A\&M University, College Station, TX 77843, USA}
\address[3]{Nuclear Solutions Institute, Texas A\&M University, College Station, TX 77843, USA}
\address[4]{Department of Physics and Geology, University of Perugia, and Instituto Nazionale di Fisica Nucleare, Secion of Perugia, Via A. Pascoli, 06123 Perugia, Italy}

\begin{abstract}
The $^{22}$Ne($\alpha$,$\gamma$)$^{26}$Mg and $^{22}$Ne($\alpha$,n)$^{25}$Mg reactions play an important role in astrophysics because they have significant influence on the neutron flux during the weak branch of the s-process. We constrain the astrophysical rates for these reactions by measuring partial $\alpha$-widths of resonances in $^{26}$Mg located in the Gamow window for the $^{22}$Ne+$\alpha$ capture. These resonances were populated using $^{22}$Ne($^6$Li,d)$^{26}$Mg and $^{22}$Ne($^7$Li,t)$^{26}$Mg reactions at energies near the Coulomb barrier. At these low energies $\alpha$-transfer reactions favor population of low spin states and the extracted partial $\alpha$-widths for the observed resonances exhibit only minor dependence on the model parameters. The astrophysical rates for both the $^{22}$Ne($\alpha$,$\gamma$)$^{26}$Mg and the $^{22}$Ne($\alpha$,n)$^{25}$Mg reactions are shown to be significantly different than the previously suggested values.
\end{abstract}

\begin{keyword}
s-process, reaction rate,  $^{22}$Ne($\alpha$,$\gamma$)$^{26}$Mg, $^{22}$Ne($\alpha$,n)$^{25}$Mg, sub-Coulomb $\alpha$-transfer reaction
\end{keyword}

\end{frontmatter}

%\linenumbers

%Introduction
\section{Introduction}

The $^{22}$Ne($\alpha$,n)$^{25}$Mg reaction is one of the two main neutron sources for the s-process - a slow neutron capture process that is responsible for the formation of about half of the elements beyond Fe \cite{2011RvMP...83..157K,2008ARA&A..46..241S}. Due to the negative Q-value (-478 keV), this reaction is activated at relatively high temperatures ($>$0.2 GK). As a result, it plays an important role in more massive stars, where higher temperatures and densities are readily available during the final phase of the core helium burning process, and the $^{22}$Ne($\alpha$,n)$^{25}$Mg reaction dominates the neutron production. These higher mass stars are expected to be the sites for the so called weak s-process, which produces isotopes with mass up to A=90.

The effectiveness of the $^{22}$Ne($\alpha$,n)$^{25}$Mg reaction as a neutron source is influenced by the $^{22}$Ne($\alpha$,$\gamma$)$^{26}$Mg radiative capture process. This reaction has a positive Q-value, which enables it to be active during the entire He-burning phase, and thus it reduces the amount of $^{22}$Ne, that is mostly produced through the $^{14}$N($\alpha$,$\gamma$)$^{18}$F($\beta^+$,$\nu$)$^{18}$O($\alpha$,$\gamma$)$^{22}$Ne reaction sequence, before the $^{22}$Ne($\alpha$,n) reaction comes into effect. Hence it is important to constrain the rates for both of these reactions.

Uncertainties for the $^{22}$Ne($\alpha$,n)$^{25}$Mg and $^{22}$Ne($\alpha$,$\gamma$)$^{26}$Mg reactions at stellar temperatures are still large and dominated by the uncertainties associated with the properties of the resonances located within the Gamow window. Several direct measurements of the excitation functions for the $^{22}$Ne($\alpha$,n)$^{25}$Mg reaction are available \cite{Giesen1993,Jaeger2001,1969NuPhA.136..481A,1993ApJ...414..735D,Harms1991}. Two resonances play particularly important role at temperatures around 0.3 GK, dominating the reaction rate. These are the resonances at 11.32 MeV and 11.17 MeV excitation energies in $^{26}$Mg. The ($\alpha$,n) strength of the 11.32 MeV resonance was obtained in several direct experiments, but the results are not consistent, ranging from $\omega\gamma_{(\alpha,n)}=83(24)$ $\mu$eV \cite{Harms1991} to $118(11)$ $\mu$eV in the most recent study \cite{Jaeger2001}, to 234(77) $\mu$eV in \cite{Giesen1993}. On the contrary, direct measurements of the ($\alpha$,$\gamma$) strength for the 11.32 MeV resonance produced consistent results. Resonance strengths of 36(4) $\mu$eV was obtained in \cite{Wolke1989} and 46(12) $\mu$eV in \cite{Hunt2019}, with a weighted average of 37(4) $\mu$eV. The situation with 11.17 MeV resonance (or resonances) is even more complicated. A resonance at 11.15 MeV  was suggested in \cite{Giesen1993}, but it was not observed in \cite{Jaeger2001} and the upper limit for its resonance strength is given instead ($<$60 neV) \cite{Jaeger2001}. It was conclusively demonstrated later that this state cannot contribute to the $\alpha$-capture reaction because it has unnatural spin-parity 1$^+$ \cite{Longland2009}. However, new resonances in the vicinity of 11.17 MeV have recently been observed \cite{Talwar2016,Massimi2017,Lotay2019} and the contribution of these new states to the $\alpha$-capture on $^{22}$Ne reaction rates is a source of uncertainty.  

There are many experiments that used indirect methods to obtain information on the properties of the levels in $^{26}$Mg which could contribute to the astrophysically important $^{22}$Ne($\alpha$,n) and $^{22}$Ne($\alpha$,$\gamma$) reactions. The resonance reaction rates of the $^{22}$Ne($\alpha$,n)$^{25}$Mg and $^{22}$Ne($\alpha$,$\gamma$)$^{26}$Mg reactions are proportional to the partial $\alpha$-widths of the resonances in $^{26}$Mg. The $^{22}$Ne($^6$Li,d) $\alpha$-transfer reaction has been used in the past to populate the levels of interest in $^{26}$Mg  \cite{Giesen1993,Talwar2016,Ugalde2007}. The most recent and very detailed work \cite{Talwar2016} utilized a $^6$Li beam of 82.3 MeV for the $^{22}$Ne($^6$Li,d) reaction, along with an $\alpha$-particle beam of 206 MeV to populate states in $^{26}$Mg using ($\alpha$,$\alpha\prime$) inelastic scattering. The authors of Ref. \cite{Talwar2016} also summarize the results of several previous studies. An extensive amount of research has been performed previously using other various techniques to obtain data on the resonance energies of $^{26}$Mg such as neutron capture studies on $^{25}$Mg (reactions such as $^{25}$Mg($n$,$\gamma$)$^{26}$Mg and $^{25}$Mg($n$,$tot$)) \cite{Massimi2017,Koehler2002,Massimi2012}, $^{26}$Mg(p,p$\prime$)$^{26}$Mg \cite{Moss1976,Adsley2018}, $^{26}$Mg(d,d')$^{26}$Mg measurements \cite{Adsley2018}, and $^{26}$Mg($\alpha$,$\alpha\prime$)$^{26}$Mg measurements \cite{Talwar2016,Adsley2017}. $^{26}$Mg($\gamma$,$\gamma\prime$)$^{26}$Mg measurements \cite{Longland2009,Schwengner2009,deBoer2010} have also been performed using polarized and unpolarized $\gamma$ rays in order to obtain information on the spin-parities of the levels of $^{26}$Mg. The $\gamma$-decaying states in $^{26}$Mg were studied recently in Ref. \cite{Lotay2019} where the excitation energies of the resonances within the Gamow window have been constrained with high precision and spin-parity assignments were suggested for some states.

It is difficult to evaluate the astrophysical importance of resonances in $^{26}$Mg observed using indirect techniques without knowledge of the spin-parities and the $\alpha$ partial widths of the populated resonances. The angular distributions of the ($^6$Li,d) reactions are not very sensitive to the transferred angular momentum. Moreover, there is a strong dependence of the spectroscopic factors and angular distributions upon the specific parameters of the optical model potentials used in the Distorted Wave Born Approximation (DWBA) analysis of the $\alpha$-transfer reactions at high energies of the $^6$Li beam ($\sim$10 MeV/A) \cite{becchetti}. The $^{26}$Mg($\alpha$, $\alpha\prime$) reaction \cite{Talwar2016} may be used to characterize states in $^{26}$Mg. However, due to high level density in $^{26}$Mg at excitation energies around 11 MeV, unique identification of states populated in different reactions is not always possible.

The present work explores the $^{22}$Ne($^6$Li,d) and $^{22}$Ne($^7$Li,t) reactions to obtain data on resonances in $^{26}$Mg in the Gamow window. Unlike previous studies, we performed these reactions at center-of-mass energies close to the Coulomb barrier. While angular distributions are even less sensitive to the transferred angular momentum at these low energies, we expected to decrease the dependence of the results on the optical potentials and to inhibit the levels that require large transferred angular momenta - the high spin states. Such states usually play a minor role in the astrophysical processes.

The $\alpha$-transfer reactions at energies close to the Coulomb barrier have been performed previously \cite{Brune1999, PhysRevC.90.042801, PhysRevLett.114.071101, PhysRevC.91.048801}. It was demonstrated that this approach produces reliable results in determining the partial $\alpha$-width for the near $\alpha$-threshold resonances \cite{PhysRevC.90.042801}.

% Experiment
\section{Experiment}

The $^{22}$Ne($^6$Li,d) and $^{22}$Ne($^7$Li,t) reactions were measured using a 1.0 MeV/u $^{22}$Ne beam delivered by the K150 cyclotron at the Texas A\&M University Cyclotron Institute. It corresponds to the $^{22}$Ne+$^6$Li center-of-mass energy of 4.7 MeV and 5.3 MeV for the $^{22}$Ne+$^7$Li, which is below the Coulomb barrier of $\sim$6 MeV. At these sub-Coulomb energies, the dominant reaction yield is at backward angles in the center-of-mass frame. The inverse kinematics provides favorable conditions for the detection of deuterons and tritons with reasonable energies of few MeV/u at small forward angles. The lithium targets were LiF of $\sim$30 $\mu$g/cm$^2$ thickness on $\sim$10 $\mu$g/cm$^2$ Carbon backing, enriched to 95$\%$ of the $^6$Li isotope, and the $^7$Li targets were made using natural Li. The energy loss of the $^{22}$Ne beam in the targets were mainly responsible for the final energy resolution of 95 keV in the deuteron and triton spectra.

We used the Multipole-Dipole-Multipole (MDM) spectrometer \cite{1986NIMPA.245..230P} to observe deuterons scattered at 5$^{\circ}$ in the lab frame. The detection, identification and tracking of light recoils (deuterons and tritons), filtered by the MDM, is provided by the modified Oxford focal plane tracking detector \cite{Spiridon2019} with the CsI(Tl) scintillator array installed at the end of Oxford detector for this experiment for better particle identification.

A silicon detector, collimated to have an opening of 0.5$^{\circ}$ was placed in the target chamber at an angle of 31$^{\circ}$ relative to the beam direction. It was used for absolute normalization, to monitor the possible target degradation, and to measure overall efficiency of the MDM spectrometer and the focal plane detector. Using the $^{22}$Ne+$^6$Li elastic scattering and also elastic scattering of 8 MeV deuteron beam on gold target, it was established that the efficiency of the setup was 87\%.

\begin{figure}[h]
\centering
  \includegraphics[width=1.0\columnwidth]{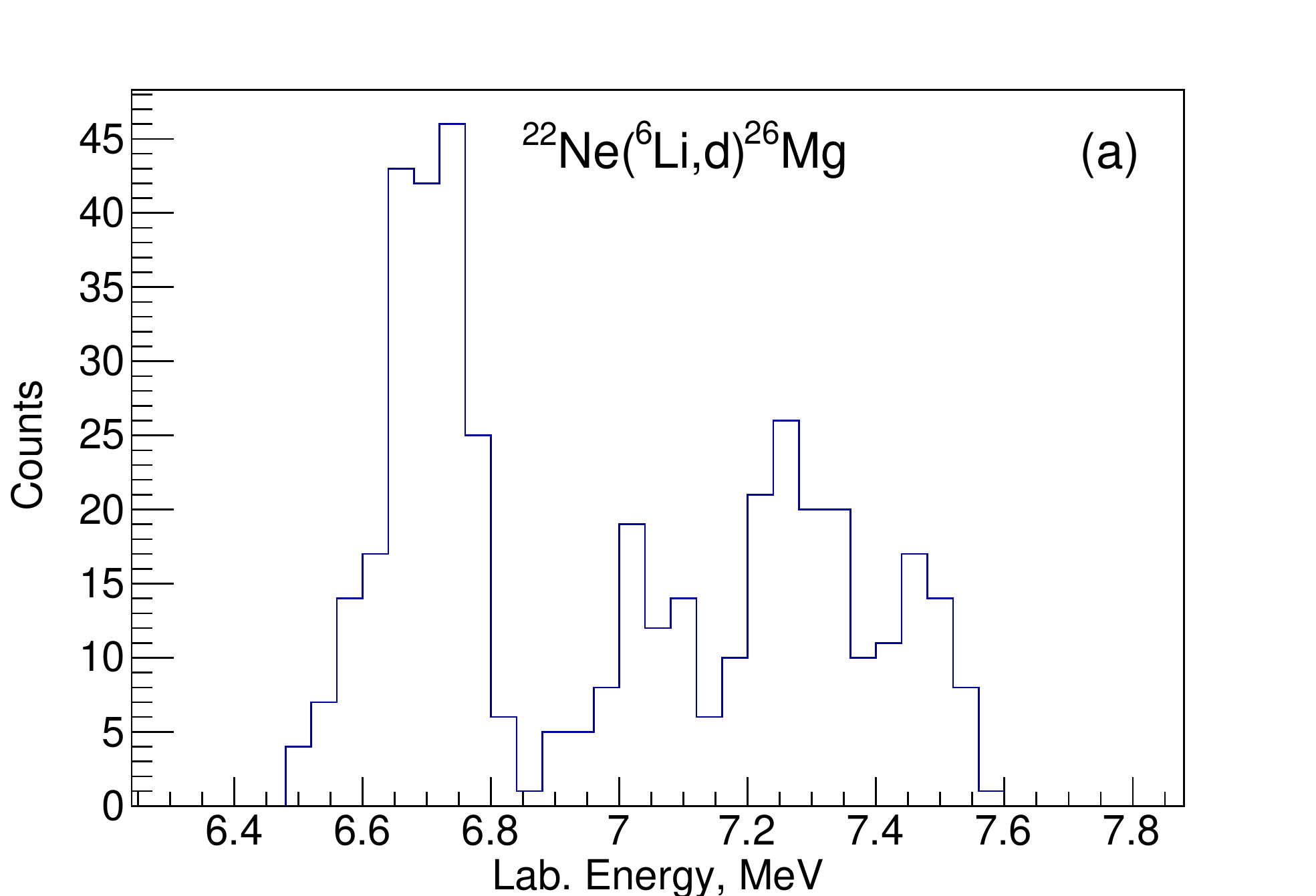}\\
  \includegraphics[width=1.0\columnwidth]{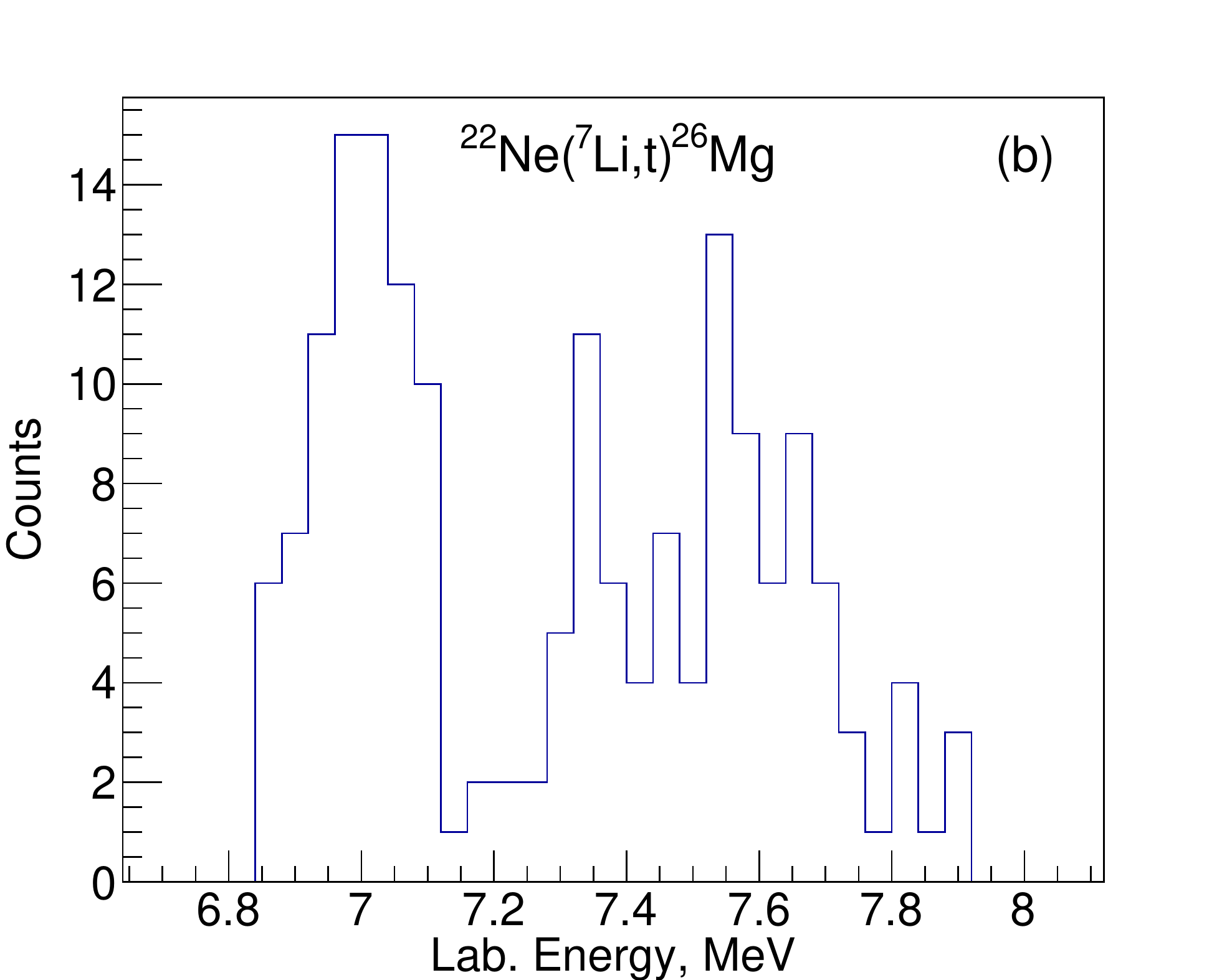}
   \caption{Energy spectrum of (a) deuterons and (b) tritons from the $^{22}$Ne($^6$Li,d)$^{26}$Mg and  $^{22}$Ne($^7$Li,t)$^{26}$Mg reactions respectively.}
   \label{fig:Ng1}
\end{figure}

\begin{figure}[h]
%[h]{0.55\textwidth}
\centering
\includegraphics[width=1.0\columnwidth]{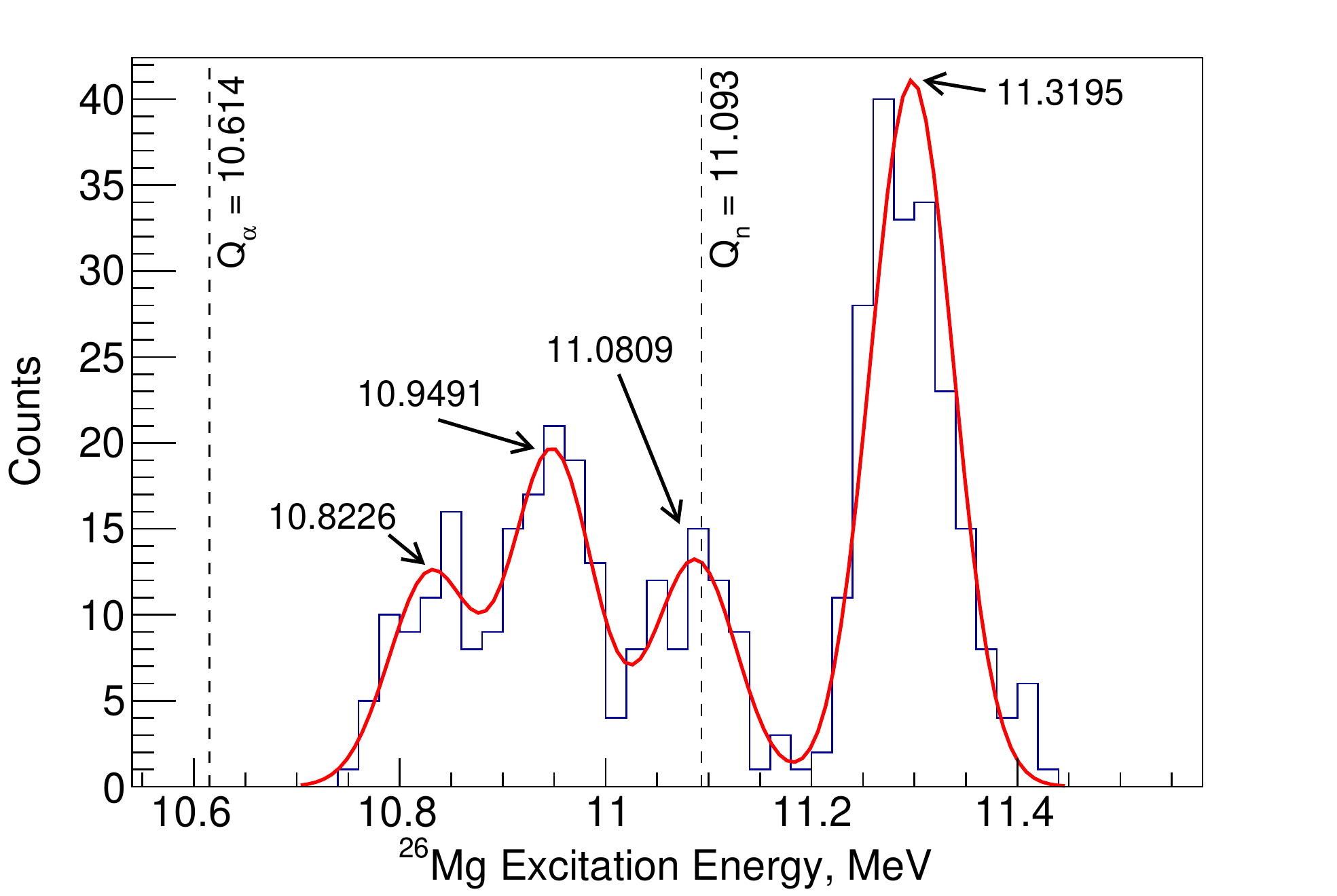}
\caption{The excitation energy spectrum of $^{26}$Mg reconstructed from the missing mass deuteron energy spectrum observed in $^{22}$Ne($^6$Li,d)$^{26}$Mg reaction. The $\alpha$ and neutron decay thresholds are shown with vertical dashed lines.}
\label{d_excitation_E}
\end{figure}

\section{Results}

Fig.~\ref{fig:Ng1} shows a deuteron energy spectrum from the $^{22}$Ne($^6$Li,d)$^{26}$Mg reaction and a triton energy spectrum from the $^{22}$Ne($^7$Li,t)$^{26}$Mg reaction measured at 5$\pm$2$^{\circ}$ lab. angle. The fields of the MDM spectrometer were set to magnetic rigidity of deuteron/triton ions with energies that correspond to population of states in the Gamow energy window for the $^{22}$Ne($\alpha$,n) and $^{22}$Ne($\alpha$,$\gamma$) reactions, between 10.7 and 11.5 MeV of $^{26}$Mg excitation. The triton missing mass spectrum from the $^{22}$Ne($^7$Li,t)$^{26}$Mg reaction was obtained with the aim of a general comparison with the higher statistics deuteron spectrum from the $^{22}$Ne($^6$Li,d)$^{26}$Mg reaction. Using the reconstructed energies and angles of the deuteron/triton particles, the Q-value of the reaction was calculated and converted to the excitation energies of $^{26}$Mg (Fig. \ref{d_excitation_E}).

Four states have been observed in the missing mass deuteron spectrum from the $^{22}$Ne($^6$Li,d)$^{26}$Mg reaction. The triton spectrum is consistent, while the counting statistics are worse due to shorter measurement. The excitation energies, center-of-mass cross sections (at 5$^{\circ}$ in the lab.) and the extracted partial $\alpha$-width (depending on the assumed spin-parity assignment) for the observed states are given in Table \ref{widths_ANC}.

\begin{table*}[p]
\caption{Excitation energies, adopted excitation energies, adopted resonances energies in center-of-mass, measured $^{22}$Ne($^6$Li,d)$^{26}$Mg cross sections and partial $\alpha$ widths for the states in $^{26}$Mg observed in this work. The widths are given for 0$^+$,1$^-$, and 2$^+$ spin-parity assignments. The preferred spin-parity assignments are boldfaced (see text). Expt. column gives the power of ten. The statistical (first) and systematic (second) uncertainties are given for cross sections and partial $\alpha$ widths.\\
{$^a$} Adopted from the most recent direct measurement of $^{22}$Ne($\alpha$,$\gamma$) by Hunt et. al \cite{Hunt2019}\\
{$^b$} The partial widths are the weighted averages between the ($^6$Li,d) and ($^7$Li,t) measurements from the present work. \\
{$^{c}$} Adopted from Lotay, et. al. \cite{Lotay2019}\\
{$^d$} Experimental cross section is normalized to $^{22}$Ne($^6$Li,$^6$Li) elastic scattering at 31$^{\circ}$ lab. (118$^{\circ}$ c.m.) which was calculated using global optical model potential given in Table \ref{long} (70 mb/sr, 70\% of Rutherford). Uncertainty associated with the specific choice of optical model potentials is included into the systematic error budget.}
\centering
\setlength{\tabcolsep}{3pt}
\begin{tabular}{c c c   c   c   c  c}
% \begin{tabular}{|c|c|c|c|c|c|c|}
\hline
         $E_{ex}$  & Adopted  $E_{ex}$ & E$_r$ & Exp. CS$^d$ & J$^{\pi}$  &  $\Gamma_\alpha$ & Expt.\\
     (MeV)    & (MeV) & (keV) & ($\mu$b/sr) & & (eV)  & \\ \hline
         
11.30(2) & 11.3195(25)$^a$& 706.6(25)$^a$ & 82 $\pm$ 6 $\substack{+13 \\ -8}$  & \bf{0$^+$}  & 6.1 $\pm$ 0.4 $\pm$ 1.0 $^b$ & -5 \\ 
                  & & & & 1$^-$  & 1.3 $\pm$ 0.1 $\pm$ 0.2 $^b$ & -5 \\ 
                  & & & & 2$^+$  & 3.0 $\pm$ 0.2 $\pm$ 0.5 $^b$ & -6 \\ %\hline

11.17 & 11.1717(30)$^{c}$ & 557.0(30)$^{c}$ & $<0.8$ & 0$^+$  & $<$3 & -9 \\ 
                  & & & & 1$^-$  & $<$6 & -10 \\ 
                  & & & & \bf{2$^+$}  & $<$1.3 & -11 \\%\hline

11.08(2) & 11.0809(40)$^{c}$ & 466.2(40)$^{c}$ & 26 $\pm$ 3 $\substack{+4 \\ -3}$  & 0$^+$  & 1.3 $\pm$ 0.1 $\pm$ 0.3 & -9 \\ 
                  & & & & 1$^-$  & 2.5 $\pm$ 0.3  $\substack{+0.7 \\ -0.5}$  & -10 \\ 
                  & & & & \bf{2$^+$}  & 5.7 $\pm$ 0.7  $\substack{+1.4 \\ -1.2}$  & -11 \\  %\hline

 10.95(2) & 10.9491(8)$^{c}$ & 334.4(8)$^{c}$ & 39 $\pm$ 4 $\substack{+6 \\ -4}$  & 0$^+$  & 1.5 $\pm$ 0.2  $\substack{+0.4 \\ -0.3}$  & -13 \\ 
  & & & & \bf{1$^-$}  & 3.0 $\pm$ 0.3 $\substack{+0.75 \\ -0.6}$ & -14 \\ 
 & & & & 2$^+$  & 6.4 $\pm$ 0.6 $\substack{+1.0 \\ -0.6}$ & -15 \\ %\hline

10.83(2) & 10.8226(30)$^{c}$ & 207.9(30)$^{c}$ & 24 $\pm$ 3 $\substack{+4 \\ -3}$  & 0$^+$  & 5.3 $\pm$ 0.7 $\substack{+1.1 \\ -1.0}$  & -21 \\ 
                  & & & & 1$^-$  & 1.0 $\pm$ 0.1 $\substack{+0.3 \\ -0.2}$ & -21\\ 
                  & & & & \bf{2$^+$}  & 2.1 $\pm$ 0.3 $\pm$0.4  & -22 \\ \hline
\end{tabular}
\label{widths_ANC}
\end{table*}

The $^{26}$Mg excitation energy spectra from both ($^6$Li,d) and ($^7$Li,t) reactions shows a similar dominance of a resonance at 11.32 MeV and serves as an indication of the dominance of the same $\alpha$-cluster transfer reaction mechanism. Out of the 4 resonances observed within the Gamow window, only the state at 11.32 MeV is above the neutron decay threshold.

The dominance of the 11.32 MeV peak within the Gamow window agrees with the most recent $^{22}$Ne($^6$Li,d)$^{26}$Mg data \cite{Talwar2016}. In contrast, there is no evidence for the 11.17 MeV resonance that was observed as an equally strong state in Ref. \cite{Talwar2016}. A peak at 11.32 MeV was also observed in $^{22}$Ne($^6$Li,d)$^{26}$Mg at 32 MeV of $^6$Li beam \cite{Giesen1993}, but the 11.17 MeV resonance is also absent. We provide a stringent upper limit for the partial $\alpha$-width of the 11.17 MeV state in this work.

The state at 11.08 MeV from the present study has been previously reported by Talwar et. al \cite{Talwar2016} at 11.085(8) MeV. The 10.95 MeV state was also present in both of the previously mentioned ($^6$Li,d) studies at 10.95 MeV in \cite{Giesen1993} and 10.951(21) MeV in \cite{Talwar2016}, as well as in Ref. \cite{Ugalde2007} at $E_x$=10.953(21) MeV.

The state at 10.83 MeV from the present study has also been seen in two previous ($^6$Li,d) studies, in Ref. \cite{Ugalde2007} at $E_x$ = 10.808(20) MeV and in Ref. \cite{Talwar2016} at 10.822(10) MeV.

\section{Analysis}

Analysis of the $\alpha$-transfer reaction cross sections was performed using Distorted Wave Born Approximation (DWBA) with code FRESCO \cite{fresco}. We used global optical potentials taken from \cite{Cook:1982zzn} for the $^{22}$Ne+$^6$Li channel and from \cite{An:2006vs} for the $^{26}$Mg+d channel (shown in Table \ref{long}). The potential parameters for the $\alpha$+d form factor were taken from Ref. \cite{KUBO1972186}. The $^{22}$Ne+$\alpha$ wave function was generated by the Woods-Saxon potential with the shape parameters given in Table \ref{long}, and the depth was fit to reproduce the binding energies of the states (see discussion below).

\begin{table*}[p]
\centering
\caption{Optical model parameters used in the  FRESCO calculations  for the $^{22}$Ne($^6$Li,d)$^{26}$Mg reaction. The radii r$_x$ are given such that $R_x = r_x {\times} A^{1/3}_T$.}
\setlength{\tabcolsep}{2pt}
 \begin{tabular}{c c c c c c c c c c c c c}
 \hline
 Reaction & V$_0$ & r$_{r}$ & a$_r$ & W$_s$ & W$_{D}$ & r$_{I}$ & a$_{I}$ & r$_{C}$ & V$_{so}$ & r$_{so}$ & a$_{so}$ & Ref. \\
 Channel & (MeV) & (fm) & (fm) & (MeV) & (MeV) & (fm) & (fm) & (fm) & (MeV) & (fm) & (fm) & \\
 \hline
 $^{22}$Ne+$^6$Li & 109.5 & 1.326 & 0.811 & 51.307 & & 1.534 & 0.884 & 1.30 & & & & \cite{Cook:1982zzn}\\
 $^{26}$Mg+d & 93.293 & 1.149 & 0.756 & 1.394 & & 1.339 & 0.559 & 1.303 & & & & \cite{An:2006vs}\\
 " & & & & & 10.687 & 1.385 & 0.715 & & 3.557 & 0.972& 1.011 &\\
 $^{22}$Ne+d & 79.5 & 1.25 & 0.8 & 10.0 & & 1.25 & 0.8 & 1.25 & 6.0 & 1.25 & 0.8 & \cite{An:2006vs}\\
 $\alpha$+d & 85.0 & 1.25 & 0.68 & & & & & 1.25 & & & & \cite{KUBO1972186} \\
 $^{22}$Ne+$\alpha$ & 138.7 & 1.23 & 0.6 & & & & & 1.25 & & & & \cite{Vlad}\\ [1ex]
 \hline
 \end{tabular}
 \label{long}
\end{table*}

To satisfy the Pauli exclusion principle, the minimum $2N+L$ values for the $\alpha$-cluster in $^{26}$Mg are 8 and 9 for positive and negative parity states respectively, where $N$ is the number of radial nodes and $L$ is the relative angular momentum of the cluster wave function. We have chosen $2N+L$=12 and 11 for positive and negative spin-parity assignment respectively, but the final partial $\alpha$ width of the states in $^{26}$Mg is insensitive to this choice. The specific shape parameters for the form factor potentials also have little influence on the partial widths. This insensitivity to the parameters of the form-factor potentials is a rather evident consequence of a peripheral nature of the $\alpha$-transfer reaction at sub-Coulomb energy. Another consequence of sub-Coulomb energy is rather weak dependence of the extracted partial width on the parameters of the optical model potentials, especially when absolute normalization is performed as a ratio to the elastic scattering cross section. 

Note that all of the $^{26}$Mg states discussed in this work are above the $\alpha$-decay threshold. Therefore, DWBA calculations of the $\alpha$-transfer to the continuum are, in principle, required. We use the bound-state approximation instead. The same approach was used in Ref. \cite{PhysRevC.90.042801} and demonstrated to work well. For a bound state, an $\alpha$-particle Asymptotic Normalization Coefficient ($C$) can be introduced. It is related to the reduced width as in Eq. \ref{width}a, where $\mu$ is a reduced mass, $R$ is a channel radius, $W$ is a Whittaker function, $S \equiv S_{\ell}(kR)$ is a shift function, and $P \equiv P_{\ell}(kR)$ is a penetrability function. Eq. \ref{width}a is evaluated at certain small binding energy between 0.1 and 1.0 MeV. Eq. \ref{width}b relates the reduced width to the partial $\alpha$ width and is evaluated at the actual center-of-mass energy of the resonance, keeping the reduced width $\gamma^2$ the same in both cases \cite{CB_private}. Partial $\alpha$-widths are calculated for several binding energies and then extrapolated linearly to the actual energy of the resonance (to negative binding energies). This extrapolation results in small width correction that does not exceed 20\%.

The partial $\alpha$ widths ($\Gamma_\alpha$) for the 4 observed resonances were calculated using the Eq. \ref{width}b. The reduced widths were evaluated by the Eq. \ref{width}a using the ANC values ($C^2$) which were determined from the ratios of the FRESCO DWBA calculations to the experimental cross sections.

\begin{subequations} \label{width}
  \begin{align}
     C^2 & = \frac{2\mu R}{\hbar^2 W^2_{-\eta,l+1/2}(2kR)} \frac{{\gamma}^2}{1+{\gamma}^2\frac{dS}{dE}} \\
      \Gamma_{\alpha} & = \frac{2 \gamma^2 P}{1 + \gamma^2 \frac{dS}{dE}}
  \end{align}
\end{subequations}

Only the 11.32 MeV resonance contributes to the $^{22}$Ne($\alpha$,n) reaction since it is neutron unbound. For this state, the width is taken as a weighted average of the ($^6$Li,d) and ($^7$Li,t) measurements. The $\Gamma_\alpha$ found using the ($^7$Li,t) measurement for the 11.32 MeV state agrees within error bars with the widths obtained for the same state using the ($^6$Li,d) measurement. The partial $\alpha$-width is largest for $J^{\pi}$ = 0$^+$ spin-parity assignment and decreases with increasing transferred angular momentum. Moreover, the resonance strength, calculated by multiplying the partial $\alpha$-width by the spin statistics factor (2J+1), is also largest for the $J^{\pi}$ = 0$^+$ spin-parity assignment. The systematic errors in Table \ref{widths_ANC} are dominated by the uncertainties associated with absolute normalization and theoretical uncertainties associated with parametrization choices for the DWBA calculations.

No more than 2 counts can be attributed to a possible state (or states) in the 11.16-11.18 MeV energy range observed in recent experiments \cite{Talwar2016,Massimi2017,Lotay2019} (see Fig. \ref{d_excitation_E}). Using the resulting experimental cross section of 0.8 $\mu$b/sr, an absolute upper limit for $\Gamma_\alpha$ of the 11.17 MeV state is calculated as 3 neV, assuming 557 keV c.m. and 0$^+$ spin-parity assignment. Adopting a tentative spin-parity of 2$^+$ \cite{Lotay2019} for this state would result in a limit of 13 peV.

The reaction rates of the $^{22}$Ne($\alpha$,n)$^{25}$Mg and $^{22}$Ne($\alpha$,$\gamma$)$^{26}$Mg reactions are proportional to the resonance strengths that are determined by the $\Gamma_\alpha$, spins and the branching ratios of the resonances in $^{26}$Mg in the Gamow window. For low energy resonances ($\Gamma_\alpha \ll \Gamma_n, \Gamma_\gamma$), the resonance strength for neutron unbound states can be written as in Eqs. \ref{strengths_ratio}, whereas for neutron bound states that contribute to the $(\alpha,\gamma)$ reaction, the resonance strength is then $\omega \gamma_{(\alpha,\gamma)}$ $\approx$ (2$J$+1) $\Gamma_\alpha$.

\begin{subequations} \label{strengths_ratio}
  \begin{align}
    \omega \gamma_{(\alpha,n)} & \approx (2J + 1) \frac{\Gamma_\alpha}{1 + \Gamma_\gamma/\Gamma_n} \label{an_eq}\\
    \omega \gamma_{(\alpha,\gamma)} & \approx (2J + 1) \frac{\Gamma_\alpha}{1 + \Gamma_n/\Gamma_\gamma} \label{ag_eq}
    \end{align}
\end{subequations}

Combining the results of this work with the new experimental data for the ($^6$Li,d) reaction obtained at energies above the Coulomb barrier \cite{Ota2019} a stringent constraint on the spin-parity assignment for the 11.32 MeV resonance can be obtained. The main result of Ref. \cite{Ota2019} is the direct measurement of the neutron to $\gamma$ branching ratio for the 11.32 MeV state - $\Gamma_n$/$\Gamma_{\gamma}$ = 1.14(26) \cite{Ota2019}. Using the weighted average between direct $^{22}$Ne($\alpha,\gamma$) measurements ($\omega\gamma_{(\alpha,\gamma)}$ = 37(4) $\mu$eV), the $\Gamma_\alpha$ of the state can be calculated using Eq. \ref{strengths_ratio}b. It is 79(13), 26(4), and 16(3) $\mu$eV for L = 0, 1 and 2, respectively. The $\Gamma_\alpha$ for the 11.32 MeV state from the present study (Table \ref{widths_ANC}) is in agreement within error bars (1.1$\sigma$) with the widths calculated from the direct $^{22}$Ne($\alpha,\gamma$) measurements but only for the $\ell$=0 case - yielding the likely 0$^+$ spin-parity assignment for the 11.32 MeV state. The $\ell$=1 assignment would produce 2.8$\sigma$ discrepancy, and the $\ell$=2 would lead to 5.0$\sigma$ discrepancy. Therefore, 0$^+$ is the highly favored spin-parity assignment, but the 1$^-$ still cannot be excluded and all other spin-parity assignments are safely excluded for the 11.32 MeV state in $^{26}$Mg.

The weighted average $\omega\gamma_{(\alpha,\gamma)}$ = 37(4) $\mu$eV for the 11.32 MeV state from previous direct measurements \cite{Wolke1989,Hunt2019} along with $\Gamma_n/\Gamma_\gamma$ = 1.14(26) from Ref. \cite{Ota2019} in Eq. \ref{strengths_ratio}, results in a neutron decay strength $\omega\gamma_{(\alpha,n)}$ = 42(11) $\mu$eV. This is within 1.7$\sigma$ of the minimum strength for this resonance obtained in Ref. \cite{Harms1991}, but certainly disagrees with all other direct measurements. If the $\Gamma_\alpha$ (for $\ell$=0) from the present measurement and the $\Gamma_n/\Gamma_\gamma$ from Ref. \cite{Ota2019} are adopted, the $\omega\gamma_{(\alpha,n)}$ would be 32(7) $\mu$eV, in good agreement with the former approach (which results in $\omega\gamma_{(\alpha,n)}$ = 42(11) $\mu$eV). However, this new ($\alpha$,n) resonance strength for the 11.32 MeV state is lower than previously reported values and results in significant reduction of the $^{22}$Ne($\alpha$,n)$^{25}$Mg reaction rate.

\begin{figure}[h]
\centering
   \includegraphics[width=1.0\columnwidth]{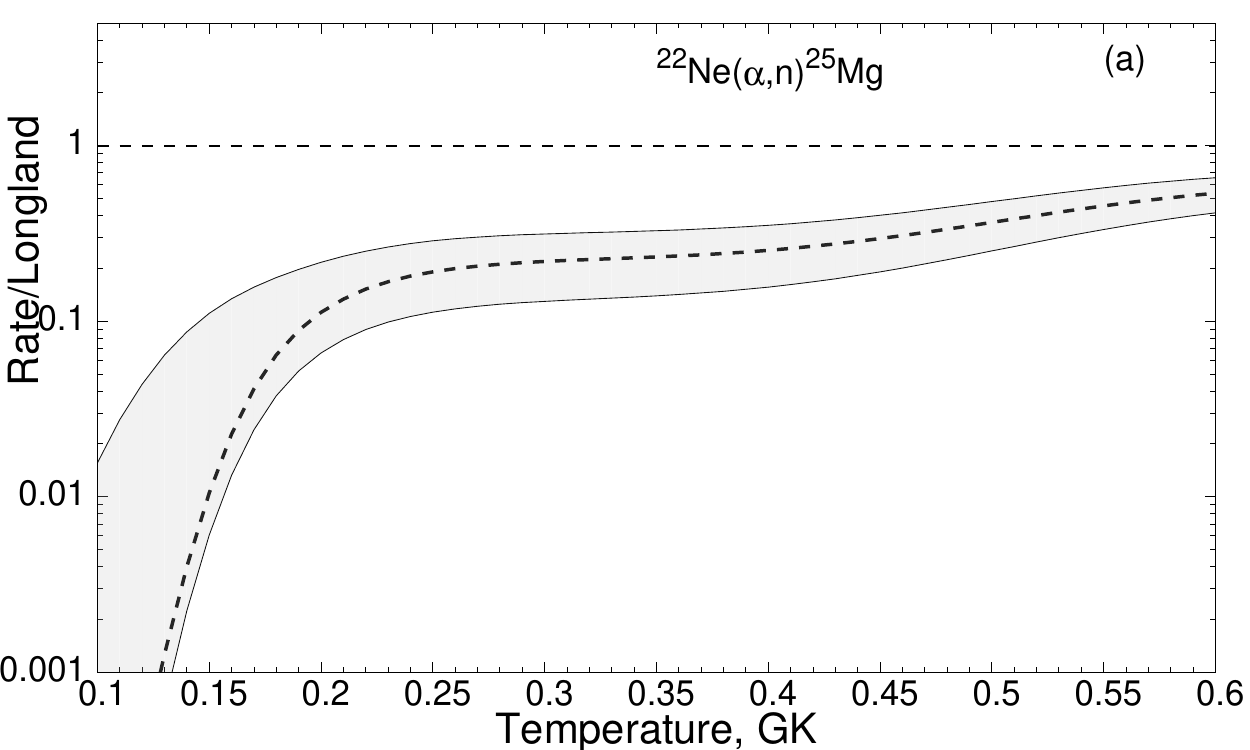}\\
      \includegraphics[width=1.0\columnwidth]{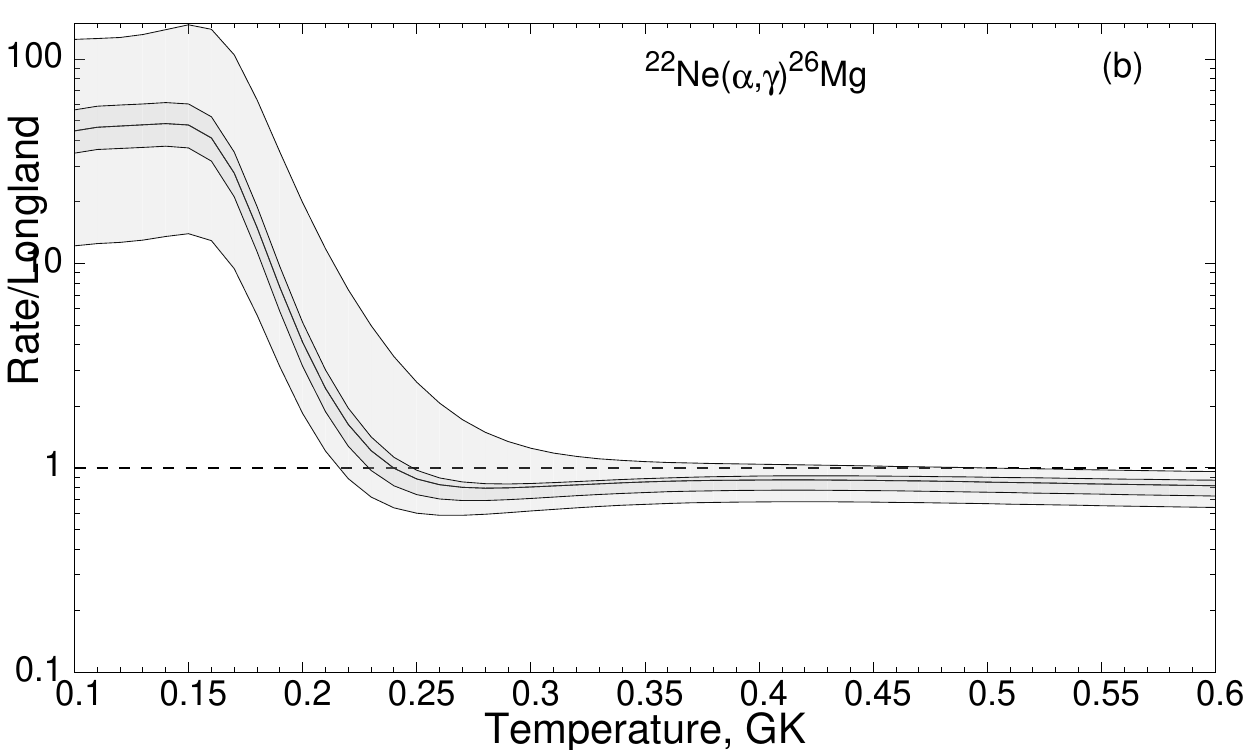}
   \caption{Ratio of the updated (a) $^{22}$Ne($\alpha$,n)$^{25}$Mg and (b) $^{22}$Ne($\alpha$,$\gamma$)$^{26}$Mg reaction rates to the recommended Monte Carlo rates of Longland, et al., \cite{Longland2012}. The light grey band represents conservative uncertainties, and the dark grey band, shown for the $^{22}$Ne($\alpha$,$\gamma$)$^{26}$Mg reaction only (b), corresponds to one $\sigma$ deviation. See text for details.}
   \label{fig:rates}
\end{figure}

The overall effect of the new constrains for the partial $\alpha$-widths of the resonances in the Gamow window on the reaction rates is demonstrated in Fig. \ref{fig:rates}. We show the ratios of the new rates to the recommended rates from Longland, et al. \cite{Longland2012}. For the $^{22}$Ne($\alpha$,n)$^{25}$Mg reaction (Fig. \ref{fig:rates}a) the dashed curve represents the rate calculated using the 11.32 MeV resonance strength of 32 $\mu$eV (for 0$^+$ assignment) and 10\% of the upper limit of the resonance strength for the 11.17 MeV resonance for the 2$^+$ assignment suggested as tentative in Ref. \cite{Lotay2019}. The strength for the higher lying states were adopted from \cite{Jaeger2001}. The conservative upper/lower limits (light grey band) correspond to 2$\sigma$ up/down deviation for the 11.32 MeV resonance strength and to the upper limit (if 0$^+$) and zero strength for the 11.17 MeV resonance respectively (assuming that 11.17 MeV resonance decays only by neutron emission). The narrow-resonance approximation was used. For the $^{22}$Ne($\alpha$,$\gamma$)$^{26}$Mg reaction we used only the states given in Table \ref{widths_ANC}. Note that the upper limit for the strength of the 11.17 MeV resonance obtained in this work is such that for the $^{22}$Ne($\alpha$,$\gamma$)$^{26}$Mg reaction it makes little difference if the state is included or not. It is still true even if we make the assumption that only $\gamma$-decay contributes to the de-excitation of this state. The conservative uncertainty band (grey region) corresponds to 2$\sigma$ deviation and simultaneously extreme assumptions for the spin-parity assignment - all four states are 0$^+$ for the upper limit and all states but 11.32 MeV are 2$^+$ states for the lower limit. The weighted average of the direct measurements was used for the ($\alpha$,$\gamma$) strength of the 11.32 MeV state - $\omega\gamma_{(\alpha,\gamma)}$=$37(4)$ $\mu$eV. It is consistent (within 1.1$\sigma$) with the value of 29(6) $\mu$eV that is obtained using Eq. \ref{strengths_ratio}b, the $\Gamma_\alpha$ measured in this work and the $\Gamma_n$/$\Gamma_{\gamma}$ ratio from \cite{Ota2019}. The dark grey band in Fig. 3(b) is a more realistic, 1$\sigma$ uncertainty with spin-parity assignments for all states except for 11.32 MeV taken from \cite{Lotay2019} - 2$^+$/1$^-$/2$^+$/0$^+$ for the 10.83/10.95/11.08/11.32 MeV states respectively. Using the data on the partial $\alpha$-widths obtained in this work it becomes possible to tightly constrain the  $^{22}$Ne($\alpha$,$\gamma$)$^{26}$Mg reaction rate, provided that the spin-parities of the resonances listed in Table \ref{widths_ANC} are reliably defined. This highlights an urgent need to firmly establish the spin-parities of the states in Table \ref{widths_ANC}. The more sophisticated Monte Carlo analysis for the reaction rates that takes into account the results of this work, includes the states observed in other studies, and also provides a comparison to the other ``recommended'' reaction rates is given in \cite{Ota2019}. It is generally consistent with the rates shown in Fig. \ref{fig:rates}, except for the low energy part of the ($\alpha$,n) rate below 0.25 GK, where the 11.112 MeV state, observed in Ref. \cite{Massimi2017}, potentially dominates the reaction rate. This resonance cannot be resolved from the 11.08 MeV state in our work, making it difficult to provide stringent limits on its strength. We do not include this state in our calculations, but one should not forget that this state may play a major role at temperatures below 0.25 GK.

\vspace{0.2cm}
\section{Conclusion}

The $^{22}$Ne($^6$Li,d)$^{26}$Mg and $^{22}$Ne($^7$Li,t)$^{26}$Mg reactions were studied with an aim to identify states in $^{26}$Mg that contribute to the $^{22}$Ne($\alpha$,n)$^{25}$Mg and $^{22}$Ne($\alpha$,$\gamma$)$^{26}$Mg reaction rates that are important nuclear physics inputs for the weak branch of the s-process. Unlike other similar studies, we explore the reaction at energies close to the Coulomb barrier, thus making the interpretation of the results less model dependent. It was confirmed that the 11.32 MeV level in $^{26}$Mg provides the dominant contribution to the $^{22}$Ne($\alpha$,n)$^{25}$Mg reaction rate at temperatures around 0.3 GK. The analysis of the data from the present work, combined with the new results of Ref. \cite{Ota2019} and previous direct measurements of the $^{22}$Ne($\alpha$,$\gamma$)$^{25}$Mg reaction showed that the most probable spin-parity assignment for this state is 0$^+$, but 1$^-$ still cannot be excluded. The $\Gamma_\alpha$ values for this state were calculated (for spin-parity assignments 0$^+$, 1$^-$ and 2$^+$). While the $\alpha$-particle reduced width of the 11.32 MeV state appears to be large, indicating importance of the $\alpha$-clustering for this $\alpha$-capture reaction, it is still significantly smaller than most direct $^{22}$Ne($\alpha$,n)$^{25}$Mg experiments indicate. Conversely, the partial $\alpha$-width for the 11.32 MeV state obtained in this work is in good agreement with the direct $^{22}$Ne($\alpha$,$\gamma$)$^{26}$Mg measurements and the $\Gamma_n$/$\Gamma_{\gamma}$ ratio obtained in Ref. \cite{Ota2019}.

The partial $\alpha$-width for three more states within the Gamow window for the $^{22}$Ne($\alpha$,$\gamma$)$^{26}$Mg reaction - 10.823 MeV, 10.949 MeV, and 11.081 MeV were obtained (assuming 0$^+$, 1$^-$, and 2$^+$ spin-parity assignments). These values provide additional constrains on the $^{22}$Ne($\alpha$,$\gamma$)$^{26}$Mg reaction rate. Moreover, no evidence for a resonance (or resonances) in the vicinity of 11.17 MeV has been observed.  As a result, a stringent upper limit for a partial $\alpha$-width of resonances in this region was obtained. This is important in the context of recent experiments, in which several natural spin-parity resonances have been observed in the vicinity of 11.17 MeV  \cite{Talwar2016,Massimi2017,Lotay2019}. Detailed discussion of implications for nuclear astrophysics will be presented elsewhere.

Another important result of this work is uncovering of evident disagreement between the results of this indirect study with the direct $^{22}$Ne($\alpha$,n)$^{25}$Mg measurements and conversely good agreement with the direct $^{22}$Ne($\alpha$,$\gamma$)$^{26}$Mg measurements and the recent branching ratio study of Ref. \cite{Ota2019}. This highlights the importance of repeating direct studies of the $^{22}$Ne($\alpha$,n)$^{26}$Mg reaction to resolve this discrepancy.

\section*{Acknowledgements}

The authors are grateful to the cyclotron team at the Cyclotron Institute for consistently reliable operation and to Carl Brune (Ohio University) for reading the manuscript and making useful comments and suggestions. This research used targets provided by the Center for Accelerator Target Science at Argonne National Laboratory.
The authors acknowledge that this material is based upon their work supported by the U.S. Department of Energy, Office of Science, Office of Nuclear Science, under Award No. DE-FG02-93ER40773, and by National Nuclear Security Administration through the Center for Excellence in Nuclear Training and University Based Research (CENTAUR) under grant No. DE-NA0003841, and the Nuclear Solutions Institute at Texas A\&M University. The authors G.V.R. and H.J. are also supported by the Welch Foundation (Grant No. A-1853). A special thanks goes out to Dr. I. Thompson, Dr. Pang Yang, Dr. T. Belyaeva, Dr. Shubchintak, and Dr. A. Moro for the valuable discussions of the FRESCO results. \newline

\bibliographystyle{elsarticle-num}
\bibliography{bib}{}

\end{document}